\documentclass[prb,twocolumn,floatfix,showpacs,preprintnumbers,groupeaddress,amsmath,amssymb,subeqn]{revtex4}
\usepackage[dvips]{graphics}

\usepackage{mathptmx}
\usepackage{psfrag}
\usepackage{graphicx}

\begin{document}

\title{Quantum-fluctuation effects on the thermopower of a single-electron 
transistor}

\author{Bj\"orn Kubala}
\author{J\"urgen K\"onig}

\affiliation{Institut f\"ur Theoretische Physik III,
Ruhr-Universit\"at Bochum, D-44780 Bochum, Germany}

\date{\today}

\begin{abstract}
We study thermal conductance and thermopower of a metallic single-electron 
transistor beyond the limit of weak tunnel coupling.
Employing both a systematic second-order perturbation expansion and a 
nonperturbative approximation scheme, we find, in addition to
sequential and cotunneling contributions, terms that are associated with
the renormalization of system parameters due to quantum fluctuations.
The latter can be identified by their logarithmic temperature dependence that
is typical for many-channel Kondo correlations. 
In particular, the temperature dependence of thermopower, which provides a
direct measure of the average energy of transported particles, reflects the 
logarithmic reduction of the Coulomb-blockade gap due to quantum fluctuations.
\end{abstract}


%
%
%

%


\maketitle

\section{Introduction}
\begin{figure}[b]
\centerline{\includegraphics[width=\columnwidth]{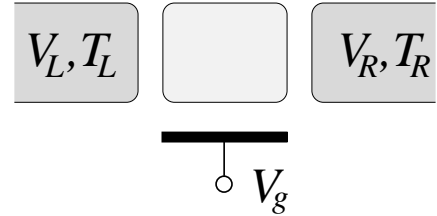}}
\caption{Setup for thermopower measurement on a single-electron transistor. The two leads are kept at different temperatures and a voltage bias $V=V_L-V_R$ is applied, such that no electrical current flows through the device. A gate voltage sets the working point and thereby the charging-energy gap $\Delta$ of the SET.}
\label{setup}
\end{figure}

Transport of electrons through a small metallic island is strongly affected by 
charging effects.\cite{FultonDolan87,KulikShekhter75,AverinLikharev86} 
Tunneling of an electron on an island with capacity $C$ is associated with an 
energy of the order of the charging energy $E_C=e^2/2C$. 
At low temperature, $k_B T\ll \Delta$, where $\Delta$ is the charging-energy 
gap between ground state and first excited charge state, transport is 
suppressed.
In a \emph{single-electron transistor} (SET), an island connected to two 
leads by tunneling junctions (see Fig.~\ref{setup}), this blockade of transport can be controlled by an additional gate, 
resulting in the well-known Coulomb oscillations of current with respect to gate voltage.\cite{Gra-Dev,Sch-Uebersicht}

If the island is well isolated from the leads, i.e., the barrier resistances $R_T^{L/R}$ are high,
\begin{equation} \label{coupling}
\alpha_0=\sum_{r=L,R}\alpha_0^r=\sum_{r=L,R}h/(4\pi^2e^2R_T^{r})\ll 1,
\end{equation}
electric transport is dominated by first-order transport in the tunnel 
conductance $\alpha_0$ (sequential tunneling).
In the Coulomb-blockade regime, where sequential tunneling is exponentially 
suppressed, inelastic cotunneling becomes important. 
In these processes of second order in $\alpha_0$, the 
energetically unfavorable charging of the island occurs only virtually.\cite{GeerligsPRL90,EilesPRL92,AverinOdintsovPhysLettA1989,AverinNazarovPRL90} But also at resonance, where sequential tunneling is present, there are 
higher-order transport contributions.
They are associated with renormalization of charging energy and tunnel 
conductance due to quantum fluctuations.
This can be qualitatively understood by mapping the SET at low temperature and
close to resonance to a many-channel Kondo problem and performing a poor 
man's scaling analysis of the latter.\cite{MatveevJETP91}
For a quantitative analysis of these quantum-fluctuation effects, a systematic
second-order perturbation expansion within a diagrammatic real-time technique 
has been performed\cite{KoenigSchoellerSchoenPRL97,KoenigSchoellerSchoenPRB98} 
and used to study different single-electron systems.\cite{TeemuPRB99,JohanssonSETquantumNoisePRL2002,KubalaPRB06}
In particular, a logarithmic reduction $\sim \alpha_0 \ln{\beta E_C}$ of the
maximum conductance, indicating a renormalization of the tunnel conductance, 
has been found, in quantitative agreement with experimental 
observations.\cite{Joyez97,Wallisser02}

In electron transport, the transfer of charge and heat are connected to
each other.
This gives rise to thermoelectric effects such as the thermal conductance and
the thermopower
\begin{equation*}
  S = \left. - \lim_{\delta T \to 0}\frac{V}{\delta T}\right|_{I=0} \,
\end{equation*}
where $V$ is the voltage due to a temperature difference $\delta T$ in the 
absence of a charge current $I$.
With the help of Onsager relations,\cite{Abrikosov} the thermopower can be 
related to the average energy $\langle \epsilon \rangle$ of the transported 
electrons relative to the Fermi energy: 
\begin{equation} \label{onsager}
  S =-\frac{\left\langle \varepsilon \right\rangle}{eT}.
\end{equation}
In macroscopic conductors, the thermopower is of the order of 
$(k_B/e) (k_BT/E_F)$, which, in general, is very small. 
This is due to the fact that the product of density of states and electron velocity squared, determining the contribution of electrons of a certain energy to the current, varies only slowly, namely
on the scale provided by the Fermi energy.
Therefore, one can perform a Sommerfeld expansion of 
$\langle \epsilon \rangle$.
The zeroth-order term of the Sommerfeld expansion vanishes.
The next-order correction, that accounts for an asymmetry of the product mentioned above 
around the Fermi energy, yields an extra factor of $k_B T/E_F$.
This is different in mesoscopic systems for which charging effects can strongly
modify the effective density of states.
In the cotunneling regime of a SET, for example, the effective density of states changes 
on the scale $\Delta$, i.e., the thermopower is of the order of 
$(k_B/e) (k_BT/\Delta)$.
An even larger thermopower is generated by sequential tunneling processes.
They are modeled by a delta-function peaked density of states around $\Delta$,
for which the Sommerfeld expansion is not applicable.
Instead, a direct evaluation of $\langle \epsilon \rangle \sim \Delta$ leads to
$S \sim (k_B/e) (\Delta / k_BT)$.
In conclusion, at the crossover from sequential to cotunneling the 
thermopower can reach values of the order of $k_B/e$.

By now a few experiments on thermal conductance or thermopower in quantum dots 
have been performed.\cite{StaringEPL93,DzurakSSC93,DzurakPRB97,MolenkampPRL05,MolenkampPRL98,MolenkampPRL99,MolenkampPRB00}
Thermoelectric effects in various mesoscopic systems have also been studied 
theoretically.\cite{BeenakkerPRB92,TurekPRB02,AndreevPRL01,AndreevPRB02,KochPRB04,BoeseEPL01,SmallPRL03,TurekPRB05,BlanterPRB97,GuttmanPRB95}
Quantum dots with discrete\cite{DzurakSSC93} (single-particle) energy level 
spectrum as well as dots closer to the metallic (quasi-continuous) 
limit\cite{StaringEPL93} have been compared to sequential-tunneling theory. 
For increased coupling and lower temperatures, systems of 
discrete level structure showed signatures of cotunneling\cite{DzurakPRB97} 
and also of Kondo physics.\cite{MolenkampPRL05} 
For the system under consideration in the present paper, a metallic 
single-electron transistor, thermopower has been investigated by taking 
into account sequential\cite{BeenakkerPRB92} and cotunneling\cite{TurekPRB02} 
processes.

In this paper, we study the effect of quantum fluctuations on the thermal
conductance and thermopower of a metallic single-electron transistor with a large number of transverse channels.
Thermopower provides a direct access to measure a renormalization of the 
charging-energy gap due to quantum fluctuations since it is closely related
to the average energy of tunneled electrons.
As we predict below, this will give rise to a logarithmic temperature 
dependence of both the slope of thermopower at resonance and of the position of the maximum of thermopower as a function of 
gate voltage.
 
This work is structured as follows. First, in Sec.~\ref{Theory}, we define 
the model Hamiltonian of the system under consideration and derive 
expressions for the thermal conductance making use of two different 
approximation schemes. 
In Sec.~\ref{Interpretation} we discuss how the thermal conductance and the
thermopower exhibit the renormalization of the 
charging-energy gap induced by quantum-fluctuations.
We summarize our results in Sec.~\ref{Conclusions}.

\section{Theory\label{Theory}}
\subsection{The system}
A metallic single-electron transistor (see Fig.~\ref{setup}) is modeled by the 
Hamiltonian
\begin{equation} \label{Hamiltonian}
        H=H_L+H_R+H_I+H_{\rm ch}+H_T=H_0+H_T.
\end{equation}
Here $H_{r}=\sum_{kn}\epsilon^{r}_{kn}a^\dagger_{{r}kn}
a_{{r}kn}$ and $H_{I}=\sum_{qn}\epsilon_{qn} c^\dagger_{qn} c_{qn}$
describe noninteracting electrons in the two leads $r=L,R$, and on
the island, respectively. 
The index $n = 1, \dots,\,N_t$ is the transverse channel index which includes the 
spin while the wave vectors $k$ and $q$ numerate the states of the electrons
within one channel.
In the following, we assume the many-channel limit $N_t \gg 1$.
Coulomb interaction of the electrons on the island is described by
the capacitance model $H_{\rm ch}=E_C(\hat{n}-n_x)^2$,
where $E_C=e^2/(2C)$ with total island capacitance $C=C_L+C_R+C_g$. 
This electrostatic energy depends on the number of excess electrons on the island, given by their number operator $\hat{n}$, as well as on applied gate and bias voltages. The latter are accounted for by the ``external charge''
$en_x=C_L V_L+C_R V_R+C_g V_g$.
To increase the number of electrons in the island from $N$ to $N+1$ one has to
overcome the charging-energy gap, the difference between neighboring charge states
 $\Delta_N = \left\langle N+1|H_{\text{ch}}
|N+1 \right\rangle -  \left\langle N|H_{\text{ch}}|N \right\rangle = E_C
\left[ 1+2\left(N-n_x\right)\right]$, which is tunable by the gate voltage via
$n_x$.
The resonance condition $\Delta_N =0$, where the charging-energy gap vanishes, 
is fulfilled at half-interger values of $n_x$.

Finally, charge transfer processes are described by 
the tunneling Hamiltonian
\begin{equation} \label{H_T}
        H_{T}=\sum_{r=L,R}\sum_{kqn} T^{rn}_{kq}
        a^\dagger_{rkn}c_{qn}{e}^{-i\hat{\varphi}}+ {\rm
        h.c.} \; .
\end{equation}
The matrix elements $T^{rn}_{kq} \equiv T^r$ are assumed to be independent of 
the states $k$ and $q$ and channel index $n$.
They determine the tunneling resistance $R_{T,r}$ of the left and 
right junction by $1/ R_{T,r} =  (2\pi e^2/ \hbar) N_t
N_{r}(0) N_I(0) |T^{r}|^2$, where $N_{I/r}(0)$ are the density 
of states of the island/leads at the Fermi level.
Note that, while the number $N_t$ of channels is large, the contribution of any one channel is so small that the total coupling remains weak and Coulomb blockade effects will occur.
The operator ${e}^{\pm i\hat{\varphi}}$ shifts the charge on the island by 
$\pm e$.
Since the left and right lead are assumed to be reservoirs with fixed 
electrochemical potential $\mu_r$ and temperature $T_r$, the phase 
$\hat \varphi$ (or its canonical conjugate, the island charge $\hat n$) is the 
only independent dynamic variable in our model. 
In general, the electron temperatures of left lead, island, and right lead can all be different from each other and differ from the lattice temperature.\cite{Kautz93}

\subsection{Conductance, thermal conductance, and thermopower}

The current $I_{r}$ flowing into reservoir $r$ can be expressed by using
correlation functions for the island charge
$C^>(t,t') = -i\langle e^{-i\hat\varphi(t)} e^{i\hat\varphi(t')} \rangle$
and $C^<(t,t') = i\langle e^{i\hat\varphi(t')} e^{-i\hat\varphi(t)} \rangle$.
For a time-translational invariant system these correlation functions
depend only on the time difference, \mbox{$C(t,t')=C(t-t')$,} and we will 
work with the Fourier transforms $C(\omega)=\int dt e^{i\omega t/\hbar} C(t)$.
The tunneling current $I=I_L=-I_R$ is determined by 
\begin{equation}\label{current}
  I_{r}=-\frac{ie}{\hbar} \int d\omega \left[ \alpha^{r+}(\omega) C^>(\omega) +
    \alpha^{r-}(\omega) C^<(\omega) \right] \, ,
\end{equation}
which includes all possible tunneling processes via the exact correlation functions $C^\gtrless (\omega)$. 
The rate functions 
\begin{equation} \label{rate}
  \alpha^{r\pm}(\omega) = \alpha_0^r \int^\infty_{-\infty} \!\!dE\, 
  f^\pm_r(E+\omega)f^\mp(E)
\end{equation}
enter, $\alpha^{r+}(\omega)$ describing tunneling of an electron from lead $r$ 
onto the island, and $\alpha^{r-}(\omega)$ from the island to lead $r$.
Here, $f^+$ denotes the Fermi function, and $f^-=1-f^+$.
Applied temperature or voltage gradients, $\delta T = T_L - T_R$ and 
$V=V_L - V_R$, are accounted for by evaluating $f_r^\pm(E+\omega)$ at 
temperature $T_r=T+\delta T_r$ and voltage $V_r$, while $f^\mp(E)$ is taken 
at $T$.
We define also 
$\alpha (\omega) = \sum_r \alpha^{r+} (\omega) + 
\alpha^{r-} (\omega)$.

The linear electrical and thermal conductances are given by
\begin{equation}
  G_V = \frac{\partial I}{\partial V}\bigg|_{V, \delta T=0}
\qquad \text{and} \qquad
  G_T =  \frac{\partial I}{\partial \delta T}\bigg|_{V, \delta T=0},  
\end{equation}
respectively.
The thermopower describes the voltage generated by a temperature difference
in the absence of an electrical current, and is related to the above mentioned
conductances by
\begin{equation*} 
S=\left.- \lim_{\delta T\to 0}\frac{V}{\delta T}\right|_{I=0} =  \frac{G_T}{G_V} \, .
\end{equation*}

To calculate the linear electrical and thermal conductance $G_V$ and $G_T$, 
we expand the rate functions up to linear order in either $V_r$ or 
$\delta T_r$
in the expression for the current, Eq.~(\ref{current}).
It is convenient to use current conservation $\sum_r I_r = 0$ to write
the current as the combination 
$I = (\alpha_0^R I_L - \alpha_0^L I_R ) / (\alpha_0^L + \alpha_0^R)$.
When expanding this combination up to linear order in either $V_r$ or 
$\delta T_r$, we immediately see that only the {\em equilibrium} correlation
functions $C^\gtrless(\omega)$, taken at $\delta T =0$ and $V=0$, enter,
since linear corrections in $V_r$ or $\delta T_r$ drop out in the combination considered.
In equilibrium, the correlation functions are related to the spectral density 
$A(\omega)$ for charge excitations on the island by 
$C^> (\omega) = - 2\pi i[1-f(\omega)] A(\omega)$ and
$C^< (\omega) = 2\pi i f(\omega) A(\omega)$.
This, eventually, leads to the linear electrical and thermal conductance
\begin{equation} 
  \label{G_V}
  G_V = G_{\text{as}}  
  \int d\omega \frac{\beta \omega/2}{\sinh \beta\omega} A(\omega) 
\end{equation}
and
\begin{equation} 
  \label{G_T}
  G_T =  - G_{\text{as}} \frac{k_B}{e}  
  \int d\omega \frac{(\beta \omega/2)^2}{\sinh \beta\omega} A(\omega) \, ,
\end{equation}
respectively,
where $G_{\text{as}}=1/\left( R^L_T + R^R_T \right)$ is the classical electrical
conductance asymptotically reached in the high-temperature limit.
In conclusion, we need to evaluate the equilibrium spectral density $A(\omega)$
to obtain the linear electrical and thermal conductance via Eqs.~(\ref{G_V})
and (\ref{G_T}).
To keep notation simple, it is convenient to introduce dimensionless
conductances $g_V$ and $g_T$, defined by
\begin{equation}
  g_V = \frac{G_V}{G_{\text{as}}}
  \qquad \text{and} \qquad
  g_T = -\frac{e}{k_B} \frac{G_T}{G_{\text{as}}} \, .
\end{equation}
As we see from Eqs.~(\ref{G_V}) and (\ref{G_T}), the dimensionless conductances
differ from each other by a factor $\beta \omega/2$ in the integrand.
This can be easily understood with the help of Eq.~(\ref{onsager}), which 
indicates that $\omega/2$ is the average energy of the lead electrons 
(measured relative to the Fermi energy) that contributes to a island charge 
excitation of the energy $\omega$.
The factor $1/2$ comes from averaging over the available phase space of 
the electronic states in the leads and the island.
Roughly speaking, on average one half of charge excitation energy comes from or
goes to the lead and the island electrons, respectively.

\subsection{Approximation schemes}

In the following we will employ two approximation schemes for calculating
the spectral density and, thus, the linear electrical and thermal conductance.
On the one hand, we will perform a systematic perturbation expansion up
to second order in the dimensionless tunnel conductance $\alpha_0$.
On the other hand, we will use a nonperturbative resummation scheme, 
the so-called ``resonant-tunneling approximation'' discussed further below.
Both schemes go beyond the weak-coupling (sequential-tunneling) limit of small 
tunnel conductances, but in different ways. 
The virtue of either scheme as compared to the other one is discussed below.
Both of these schemes are based on a real-time diagrammatic technique 
introduced in Ref.~\onlinecite{SchoellerSchon94}. Here, we will make use of 
known results of these methods without the need for an explicit recalculation of 
the diagrams. Therefore, in this paper, we will not discuss rules for 
constructing and evaluating diagrams, but refer the interested reader to the 
existing literature.

\subsubsection{Systematic perturbation expansion}\label{subsection expansion}

We perform a systematic perturbation expansion of the correlation functions
$C^\gtrless (\omega)=\sum_{k=0}^\infty C^{\gtrless (k)}(\omega)$ and, 
therefore, automatically for the spectral density 
$A(\omega)=[C^<(\omega) - C^>(\omega)]/(2\pi i) = \sum_{k=0}^\infty A^{(k)}
(\omega)$, where the index $k$ denotes the power of $\alpha_0$ 
in the expansion.
The real-time method yields diagrammatic representations of the correlation 
functions in different order (see Fig.~3 of 
Ref.~\onlinecite{KoenigSchoellerSchoenPRB98}), which are calculated as sketched
in Sec.~III of Ref.~\onlinecite{KoenigSchoellerSchoenPRB98}.

To \emph{lowest order}, the spectral density needed is simply
\begin{equation}
  A^{\text{seq}} (\omega) = \sum_N \left( P_N + P_{N+1} \right) \delta(\omega - \Delta_N)
\end{equation}
with the equilibrium probabilities (to zeroth-order in $\alpha_0$)
\begin{equation*}
  P_N = \exp{[- \beta E_{\text{ch}}(N)]}/Z 
  \quad \text{with} \quad 
  Z=\sum_N \exp{[- \beta E_{\text{ch}}(N)]} \, ,
\end{equation*}
to find the island in charge state $N$.
As a result, the dimensionless linear electrical and thermal conductances are
\begin{eqnarray}
  g_V^{\text{seq}} &=& \sum_N \left( P_N + P_{N+1} \right)
  \frac{\beta\Delta_N/2}{\sinh{ \beta\Delta_N }}
\label{G_V_seq1}
\\
  g_T^{\text{seq}} &=& \sum_N \left( P_N + P_{N+1} \right)
  \frac{(\beta\Delta_N/2)^2}{\sinh{ \beta\Delta_N }} \, .
\label{G_T_seq1}
\end{eqnarray}
For low temperatures at most two charge states contribute, e.g., for $n_x \approx
0.5$ only the term $N=0$ enters, and $P_0 + P_1=1$.
Since, in the lowest order, the only allowed charge excitation energies are $\Delta_N$, the average
energy of the contributing electrons for transitions between charge state
$N$ and $N+1$ is $\Delta_N/2$.

For the \emph{next-order} contribution, we use correlation functions in the limit 
of vanishing applied voltage and temperature bias from  Ref.~\onlinecite{KoenigSchoellerSchoenPRB98}, namely Eqs.~(12),(14), and (15) together 
with Eqs.~(20) and (60) there.
We can perform all integrals in Eq.~(\ref{G_T}) above analytically to find the 
complete second-order contribution to the linear electrical and thermal 
conductances as a sum of four terms, 
\begin{equation} \label{GT2}
  g_{V/T}^{(2)} =
  g_{V/T}^{\text{cot}} + g_{V/T}^{\tilde\alpha} + g_{V/T}^{\tilde\Delta} + 
  g_{V/T}^{2e}
  \; .
\end{equation}
\addtocounter{equation}{-1}
For the dimensionless thermal conductance, we get the lengthy but complete
expressions
\begin{widetext}
\begin{subequations}
\begin{eqnarray}
  \label{GT2_1}  
  g^{\text{cot}}_{T} &=& \sum_N P_N \Bigg[ \;
    a_{N-1} \Delta_{N-1} \partial^2 \phi_{N-1} + 
    a_N \Delta_N \partial^2 \phi_N 
    + \frac{a_N + a_{N-1}}{2} \cdot \frac{\phi_N -\phi_{N-1} + 
    \Delta_{N-1} \partial \phi_{N-1} - \Delta_N \partial \phi_N }{E_C} 
    \Bigg]
  \\ 
  \label{GT2_2}
  g^{\tilde\alpha}_{T} &=& 
  \sum_N \, a_N \frac{\beta \Delta_N/2}{\sinh \beta\Delta_N} 
  \left( P_N + P_{N+1} \right) \Biggl[ 
    \partial \left( 2\phi_N + \phi_{N-1} + \phi_{N+1} \right) 
    +\frac{\phi_{N-1} - \phi_{N+1}}{E_C}
  \\ \nonumber
  &&\phantom{\!\!\!\times\Biggl[}\;\;\;  
    + \beta \sum_{N'} P_{N'} \left( \phi_{N'-1} - \phi_{N'} \right)
    - \frac{\beta P_N \left( \phi_{N-1}-\phi_N \right)
      + \beta P_{N+1} \left( \phi_N -\phi_{N+1} \right)}{P_N + P_{N+1}}  
    \Biggr]
  \\ 
  \label{GT2_3}
  g^{\tilde\Delta}_{T} &=& \sum_N\;   
  \partial \left[ a_N \frac{\beta \Delta_N/2}{\sinh \beta\Delta_N} \right]
  \left( P_N + P_{N+1} \right)   
  \left( 2\phi_N - \phi_{N-1} - \phi_{N+1} \right) 
  \\ 
  \label{GT2_4}
  g^{2e}_{T} &=& \sum_N\; \frac{a_N+a_{N-1}}{2} \cdot  
  \frac{\beta (\Delta_N+\Delta_{N-1})}
       {\sinh \beta(\Delta_N+\Delta_{N-1})} \left( P_{N-1} + P_{N+1} \right) 
       \left[ \frac{\phi_N - \phi_{N-1}}{2 E_C}  -\frac{  \Delta_{N-1} 
	   \partial \phi_N + \Delta_N \partial \phi_{N-1} }{\Delta_{N}+
	   \Delta_{N-1}  } \right]
       .
\end{eqnarray}
\end{subequations}
\end{widetext}
Here, $\partial$ stands for $\partial/\partial \Delta_N = -(1/ 2E_C)
(\partial / \partial n_x)$ and we used the definition 
\mbox{$\phi_{N}=(\alpha_0^L + \alpha_0^R ) \Delta_N \, {\rm Re} \, 
\Psi \left(i \beta \Delta_N /2\pi\right)$,} where $\Psi$ denotes the digamma 
function.
Furthermore, we defined $a_N \equiv \beta \Delta_N /2$.
The result for the dimensionless electrical conductance $g_V$ is the same but 
without the factors $a_N$, $a_{N-1}$ and $(a_N+a_{N-1})/2$ in accordance with 
Ref.~\onlinecite{KoenigSchoellerSchoenPRB98}.
The factors $a_N$ and $a_{N-1}$ account for the average energy of the
lead electrons contributing to transport.

For the interpretation of the four terms we follow the reasoning put forward
in Ref.~\onlinecite{KoenigSchoellerSchoenPRB98}.
The first term $g^{\text{cot}}_{T}$ models cotunneling processes, where an
electron is transferred through the whole device without changing the charge of 
the island.
This is the dominant transport contribution far from the resonance of 
sequential tunneling.
We can identify this term with the ``reguralized'' cotunneling result postulated in Ref.~\onlinecite{TurekPRB02}.
In fact Eq.~(\ref{GT2_1}) stems from the integral expression
\begin{align*}
 g^{\text{cot}}_{T} &= \sum_N \;P_N \alpha_0 \int\! d\omega \;
\omega 
\frac{(\beta\omega/2)^2}{2\sinh^2(\beta\omega/2)}
\tag{\ref{GT2_1}'} \label{GT2_1'}\\
& \phantom{-\frac{k_B}{e} G_{\text{as}}\;}\times {\text{Re}}\left({1\over \omega-\Delta_N+i0^+}-{1\over\omega-\Delta_{N-1}+i0^+}\right)^2 ,
\end{align*}
which yields Eq.~($30$) of Ref.~\onlinecite{TurekPRB02} in the regime 
considered there. 
Note that the infinitesimal imaginary parts of the denominator arises naturally
within the diagrammatic theory, not requiring regularization by hand as in  
Ref.~\onlinecite{TurekPRB02}.
At low temperature and away from resonance, e.g., in the Coulomb-blockade 
valley with $P_0=1$, we can make use of the expansion 
${\text{Re}} \, \Psi(ix) = \ln |x| + 1/(12x^2) + 1/(120x^4) + \ldots $
to get
\begin{eqnarray}
  \label{cot_V}
  g_V^{\text{cot}} &=& \alpha_0 \frac{2\pi^2}{3} (k_BT)^2 
  \left( \frac{1}{\Delta_0} - \frac{1}{\Delta_{-1}} \right)^2
\\
  \label{cot_T}
  g_T^{\text{cot}} &=& \alpha_0 \frac{8\pi^4}{15} (k_BT)^3 
  \left( \frac{1}{\Delta_0} - \frac{1}{\Delta_{-1}} \right)^2
  \left( \frac{1}{\Delta_0} + \frac{1}{\Delta_{-1}} \right)
\end{eqnarray}
in accordance with Ref.~\onlinecite{TurekPRB02}.

Away from resonance, $g_{V/T}^{\text{cot}}$ is the only second-order 
contribution.
When approaching the resonance, two more terms, $g_{V/T}^{\tilde \alpha}$ and 
$g_{V/T}^{\tilde\Delta}$ come into play.
They are associated with sequential-tunneling processes but with renormalized parameters: 
$g_{V/T}^{\tilde \alpha}$ is the first correction term to sequential tunneling due to renormalization of the  
tunnel-coupling strength, $g_{V/T}^{\tilde\Delta}$ the respective correction due to a renormalized 
charging-energy gap.
The relation of these terms to renormalization is discussed in more detail
below.
At low temperature, only the term $N=0$ contributes and $P_0+P_1=1$, so that
Eqs.~(\ref{GT2_2}) and (\ref{GT2_3}) reduce to
\begin{align*}
  g^{\tilde \alpha}_{T} &= 
\frac{(\beta \Delta_0/2)^2}{\sinh \beta\Delta_0}\left[  \partial \left( 2\phi_0 + \phi_{-1} + \phi_{1} \right) + \frac{\phi_{-1} - \phi_{1}}{ E_C} \right] 
 \tag{\ref{GT2_2}'}\label{GT2_2'} \\ \tag{\ref{GT2_3}'} \label{GT2_3'}
  g^{\tilde\Delta}_{T} &= \partial \left[   \frac{(\beta \Delta_0/2)^2}{\sinh \beta\Delta_0} \right] \left( 2\phi_0 - \phi_{-1} - \phi_{1} \right) \, .
\end{align*}

The fourth term, $g^{2e}_{T}$ describes cotunneling processes in which the 
charge of the island is changed by $2e$.
Since the total change of the charging energy between charge state
$N+1$ and $N-1$ is $\Delta_N+\Delta_{N-1}$, the factor $(a_N+a_{N-1})/2$
accounts for the average energy per contributing lead electron.
To overcome the charging energy for two electrons, a large temperature is 
required.
The term $g^{2e}_{T}$ vanishes at low temperature, and hence will not 
be of importance in the following.

The virtue of the perturbation expansion lies in the fact that (i) all 
second-order contributions are systematically taken into account, (ii)
their identification with cotunneling processes and renormalization 
corrections to sequential tunneling is straightforward, and (iii) all 
expressions are unambiguously fixed by the system parameters without any
remaining cutoff parameters.
With increasing tunnel-coupling strength or lowering temperature, however, 
the second-order perturbation theory will become insufficient.
Therefore, we also apply a different approximation scheme as described in the
the following.

\subsubsection{Resonant-tunneling approximation}

The so-called resonant tunneling approximation (RTA) has been introduced in 
Ref.~\onlinecite{SchoellerSchon94} as a nonperturbative treatment of quantum 
fluctuations.
It amounts to resummation of a certain diagram class, including contributions
of arbitrary high order in the tunnel-coupling strength. 
In particular, only two charge states $N=0,1$, and only
density-matrix elements that are at most twofold off-diagonal are taken into 
account.
For details of the derivation, we refer to Ref.~\onlinecite{SchoellerSchon94}. 

Within RTA, the equilibrium spectral function is found\cite{SchoellerSchon94} 
to be 
\begin{equation}
  A(\omega) = \frac{\alpha(\omega)}{|\omega - \Delta_0 -\sigma(\omega)|^2}
\label{spec_RTA}
\end{equation}
with the self-energy 
\begin{equation*}
 \sigma(\omega) = \int_{-\infty}^{\infty} d\omega' \frac{\alpha(\omega)}{ \omega - \omega' + i0^+} \, .
\end{equation*}
Real and imaginary part are given by
\begin{eqnarray*}
  {\rm Re}\, \sigma(\omega) &=& -2 \alpha_0 \omega
  \left[ \ln \left( \frac{\beta D}{2\pi} \right) - \;{\rm Re} \, \Psi 
    \left(i \frac{ \beta \omega}{ 2\pi} \right) \right]
  \\
  \label{Im sigma}
	{\rm Im}\, \sigma(\omega) &=& -\pi \alpha(\omega) \, ,
\end{eqnarray*}
where $D$ is a high-energy cutoff of the order of the charging energy or band
width.
The expression for the electrical and thermal linear conductance follows from 
Eqs.~(\ref{G_V}) and (\ref{G_T}).

The virtue of the RTA is that, due to the resummation of higher-order 
contributions, lower temperature and higher values of the tunnel-coupling 
strength are accessible.
This is also indicated by the fact that the self-energy $\sigma(\omega)$, 
describing renormalization of the charging energy-gap and the tunnel-coupling 
strength, appears in the denominator of Eq.~(\ref{spec_RTA}).
On the other hand, the truncation of the Hilbert space to two charge states
leaves a high-energy-cutoff dependence of the results.
Thus RTA is suited for describing effects of the qualitative temperature 
dependence due to quantum fluctuations at low temperature.
For quantitative results at higher temperature, the systematic second-order
perturbation expansion is more reliable.

\subsection{Renormalization effects}

The main result of this paper is the appearance of renormalization 
effects in thermoelectric quantities.
Therefore, we comment in this subsection on the relation between 
quantum-fluctuation induced renormalization and the electrical and thermal 
conductance in more detail.
This discussion is equally valid for the results of the electrical and
thermal conductance, and follows along the line of 
Ref.~\onlinecite{KoenigSchoellerSchoenPRB98}.

The notion of system-parameter renormalization is the central idea of
all renormalization-group (RG) schemes.
An effective low-energy model is derived by successively integrating out 
high-energy degrees of freedom in the leads.
A poor man's scaling version of such an RG scheme for the two-charge-state
approximation of the metallic SET has been performed in 
Ref.~\onlinecite{MatveevJETP91} by mapping it to a many-channel Kondo model.
During the RG procedure, both the tunnel coupling strength $\alpha_0$ and
the charging-energy gap $\Delta_0$ becomes renormalized as a consequence of
the tunnel coupling between island and lead electrons.
The renormalized values $\tilde \alpha$ and $\tilde \Delta$ are, within this
poor man's scaling scheme, 
\begin{equation} \label{renormscaling}
  \frac{\tilde{\alpha}}{\alpha_0} = \frac{\tilde{\Delta}}{\Delta_0}
  = \frac{1}{ 1+2\alpha_0 \ln ( D/\omega_C)},
\end{equation} 
where $D$ is the high-energy cutoff (the smaller of charging energy or band 
width) and $\omega_C$ the low-energy scale at which the RG procedure stops 
(here the larger of temperature $k_BT$ or charging-energy gap $\Delta_0$).
As a consequence of the large number of transverse channels in the tunnel
contacts, the charging-energy gap and the tunnel-coupling strength
are renormalized towards lower values, with a logarithmic dependence on the
high- and low-energy cutoff.
While the result is inherently nonperturbative (an expansion of the 
denominator yields all powers of $\alpha_0$), it is rather qualitative as the 
numerical coefficient $D/\omega_C$ is unknown.

In the spirit of an RG picture, an effective low-energy theory of transport
that takes into account renormalization is obtained by taking the 
sequential-tunneling formula but with renormalized system parameters
$\tilde \alpha$ and $\tilde \Delta$ instead of the bare ones $\alpha_0$ and 
$\Delta_0$.
This amounts to 
\begin{equation} \label{renorm}
  G (\alpha_0,\Delta_0) = G^{\text{seq}}(\tilde\alpha,\tilde\Delta) + 
  \text{regular terms} \, ,
\end{equation}
for the electrical or thermal conductance.
The ``regular terms'' represent higher-order contributions, such as cotunneling
processes, that are not associated with renormalization.
The latter are not included in the RG procedure, and are not considered in the 
following.

To relate the second-order transport contributions to renormalization, we 
expand Eq.~(\ref{renorm}) up to second order in $\alpha_0$,
\begin{equation*}
  G^{\text{seq}}(\tilde\alpha,\tilde\Delta) =  
  \frac{\tilde\alpha}{\alpha_0}  
  G^{\text{seq}}(\alpha_0,\Delta_0) +
  \left( \tilde\Delta - \Delta_0 \right) 
  \frac{ \partial G^{\text{seq}}(\alpha_0,\Delta_0) }{\partial \Delta_0} \, .
\end{equation*}
By comparison with Eqs.~(\ref{GT2_2'}) and (\ref{GT2_3'}), we obtain
\begin{eqnarray}
\frac{\tilde{\alpha}}{\alpha_0}  \!\!&=&\!\! 1  - 2 \alpha_0 
  \left\{ -1+\ln \left( \frac{\beta E_C}{\pi} \right) -\partial_{\Delta_0} 
  \left[ \Delta_0 \, {\rm Re} \, \Psi \left( 
    i\frac{ \beta \Delta_0}{ 2\pi } \right) \right] \right\}
  \nonumber \\
  && \label{alpharenormcot}\\
  \frac{\tilde{\Delta}}{ \Delta_0}  \!\!&=&\!\! 1 - 2 \alpha_0
  \left[ 1 + \ln \left( \frac{\beta E_C}{\pi}
    \right) - {\rm Re} \, \Psi \left( i\frac{ \beta \Delta_0}{2\pi } 
    \right)\right] \, . 
\label{Deltarenormcot}
\end{eqnarray}
Within the RTA we find\cite{SchoellerSchon94}
\begin{equation} \label{renorm_RTA}
\frac{\tilde{\alpha}}{\alpha_0} = \frac{\tilde\Delta}{\Delta_0} = 
\frac{1}{1 + 2 \alpha_0 \left[ \ln \left( \frac{\beta D}{2\pi} \right) 
      - {\rm Re} \, \Psi \left( i \frac{\beta \tilde{\Delta} }{2\pi} 
      \right) \right] } \, .
\end{equation}

The RTA result is nonperturbative in $\alpha_0$, and it resembles the 
structure of the poor man's scaling RG result in Eq.~(\ref{renormscaling}).
Its numerical value remains undetermined as the exact form of the
high-energy cutoff $D$ is left unspecified.
This contrasts to the result from second-order perturbation theory.
There, all numerical constants are specified.
On the other hand, the renormalization is determined only up to linear 
corrections in $\alpha_0$. 
Indeed, this correction can be considered as the lowest-order term of an expansion in 
$\alpha_0 \ln{ ( E_C / \max{ \{ \Delta_0, k_BT \} } ) }$ [cf. Eq.~(\ref{renorm_RTA})], which serves as small parameter. 
We conclude with the remark that the interpretation of some of the second-order
contributions as first-order transport but with renormalized parameters 
was recently supported by analyzing the full counting statistics of 
electrical transport through a metallic SET.\cite{BraggioKoenigFazio06}
There, the functional dependence of the cumulant generating function on the
counting fields enables an unambiguous identification of sequential and
cotunneling, in full support of our interpretation above.

\section{Results\label{Interpretation}}

\subsection{Thermal conductance}

\begin{figure}[]
\centerline{\includegraphics[width=0.95\columnwidth]{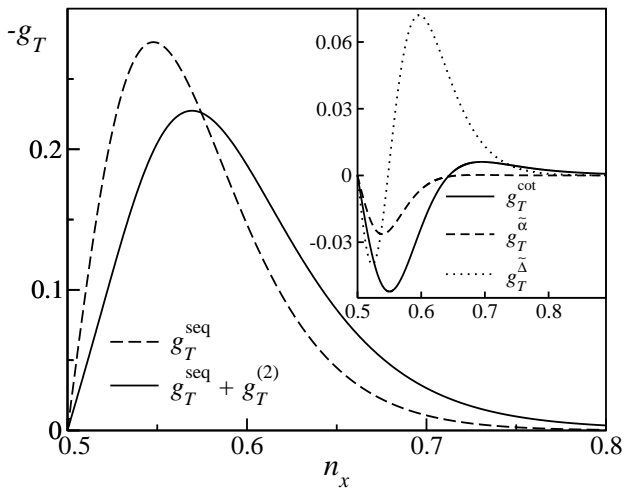}}
\caption{Thermal conductance of an SET calculated to first and second order in tunneling coupling for $k_B T/E_C = 0.05$ and $\alpha_0 = 0.04\,$. The inset shows the different contributions in second order, due to cotunneling ($g^{\text{cot}}_{T}$), renormalization of the tunnel-coupling strength ($g^{\tilde \alpha}_{T}$) and of the charging-energy gap ($g^{\tilde \Delta}_{T}$). Renormalization leads to a broadening and a suppression of the thermal conductance as compared to the sequential-tunneling result and cotunneling yields an algebraically decaying contribution in the Coulomb-blockade valley. 
}
\label{G_Tfig}
\end{figure}
Figure~\ref{G_Tfig} shows the first- and second-order contributions to the 
thermal conductance, i.e., the sequential-tunneling result, 
Eq.~(\ref{G_T_seq1}), and the different contributions to second order,
Eq.~(\ref{GT2}), in the inset. 
For the low temperature considered here, $g^{2e}_{T}$ vanishes, as it 
corresponds to two electron processes---leaving the dot in a state, where 
the electrostatic energy is changed by more than $E_C$.
As discussed above, $g^{\text{cot}}_T$ reproduces the regularized cotunneling
result; it is the dominating contribution away from resonance 
($\beta \Delta_0 \gg 1$), as it decays algebraically only. 
From Eqs.~(\ref{cot_V}) and (\ref{cot_T}), we see that the cotunneling
contribution to the electrical and thermal conductance scales with
$(T/\Delta_0)^2$ and $(T/\Delta_0)^3$, respectively.
The $(T/\Delta_0)^2$ behavior of the electrical conductance is understood from 
the fact that each tunneling rate contributes a factor $T$ while the 
$\Delta_0^2$ denominator is that of standard second-order perturbation theory.
At fist glance, one might expect a $T/\Delta_0$ behavior for the thermal 
conductance due to the relative factors $a_N = \beta\Delta_N/2$ in Eq.~(\ref{GT2_1}).
However, it turns out that terms in the thermal conductance stemming from the 
lowest order in $1/(\beta \Delta_0)$ in the expansion of $\phi_0$  cancel out 
when expanding Eq.~(\ref{GT2_1}).
Since the expansion of $\Psi(ix)$ has only even powers in $x$, the first
nonvanishing contribution to the thermal conductance scales with
$(T/\Delta_0)^3$.

At resonance, the terms $g^{\tilde \alpha}_{T}$ and $g^{\tilde \Delta}_{T}$,
associated with renormalization of the tunnel coupling $\tilde \alpha$ and the 
charging-energy gap $\tilde \Delta$, become important.
The renormalization of the tunnel coupling strength towards lower value
results in a reduction of the peak height.
The renormalization of the charging-energy gap shifts the system effectively 
closer to resonance and consequently yields a broadening of the resonance 
structure. 
In other words, the renormalization of coupling is reflected in the suppression
of the maximum value of thermal conductance, the renormalization of the 
charging-energy gap in the shift of the maximum's position.

The results for thermal conductance look rather familiar from the conductance results and do not clearly showcase unexpected features. Looking at the thermopower, however, we can gain new and interesting insights in the mechanisms of electron transport through our system and how it is influenced by quantum fluctuations. This is owed to the intuitively appealing interpretation of thermopower as measure of the average energy of transported particles, see Eq.~(\ref{onsager}).

\subsection{Thermopower}

\begin{figure}[]
\centerline{\includegraphics[width=0.95\columnwidth]{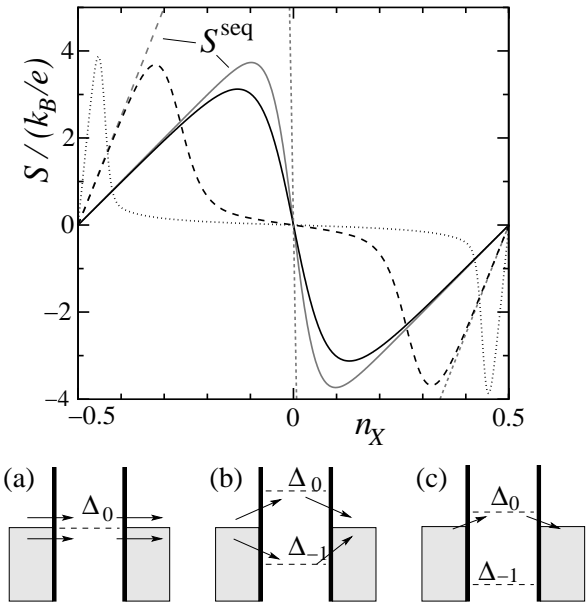}}
\caption{Thermopower within perturbative calculation for  $\alpha_0 =0.002$
and $k_B T/E_C = 0.1$ (solid line), $0.04$ (dashed), and $0.01$ (dotted).
}
\label{SdivT}
\end{figure}

The thermopower as a function of the gate charge $n_x$ for different 
temperatures is displayed in Fig.~\ref{SdivT}.
We show the full result (black lines)
\begin{equation} \label{Squotient}
  S = \frac{ G_T^{\text{seq}} + G_T^{(2)} }{ G_V^{\text{seq}} + G_V^{(2)} }\, ,
\end{equation}
that takes into account all first- and second-order contributions to the 
electrical and thermal conductance, and, for comparison, also the pure
sequential-tunneling result $S^{\text{seq}}=G_T^{\text{seq}}/G_V^{\text{seq}}$
(gray lines).
The thermopower vanishes at both integer and half-integer values of $n_x$.
At resonance, i.e., at half-integer values of $n_x$, the thermal conductance
vanishes due to a cancellation of transport contributions from lead electrons 
above and below the Fermi level, that generate {\em the same} charge 
excitation [Fig.~\ref{SdivT}(a)].
In the middle of the Coulomb-blockade valley, i.e., at integer values of $n_x$,
the zero is due to a cancellation of processes that involve adjacent charge
excitations [see Fig.~\ref{SdivT}(b)].
In between [situation sketched in Fig.~\ref{SdivT}(c)], the thermal conductance and, thus,
the thermopower is finite, with alternating sign at integer and half-integer
values of $n_x$.

At larger temperature (solid line in  Fig.~\ref{SdivT}), sequential tunneling 
dominates.
Results for this regime have first been derived in Ref.~\onlinecite{BeenakkerPRB92}. 
We recover a sawtooth behavior, with a linear increase as long as transport 
predominantly involves only one transition $N \leftrightarrow N+1$ of 
charge states.
Then the average energy of transported particles and correspondingly the 
thermopower is proportional to $\Delta_N/2$.
Around $n_x \approx N$, the adjacent transition $N-1 \leftrightarrow N$, that 
contributes with an opposite sign, comes into play.
This gives rise to a sharply falling edge of the sawtooth with a width given 
by temperature. 

At lower temperatures (dashed and dotted lines), sequential tunneling dominates 
transport only around half-integer values of $n_x$, but cotunneling takes
over in the Coulomb-blockade valley in between, resulting in a suppression of the rising edge of the sawtooth. Instead thermopower decays with  
$T/\Delta$ away from resonance, as seen from Eqs.~(\ref{cot_V}) and (\ref{cot_T}).

\begin{figure}[b]
\centerline{\includegraphics[width=\columnwidth]{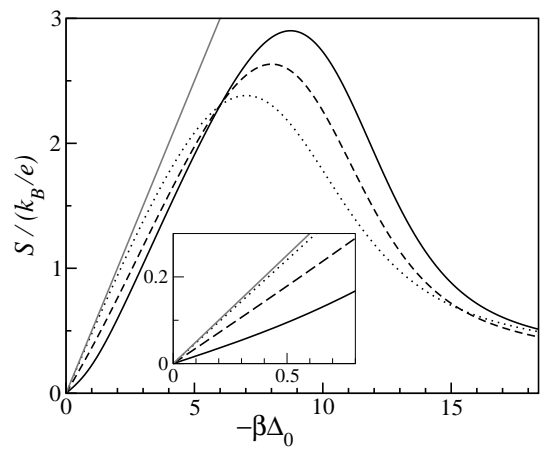}}
\caption{Thermopower as a function of $\beta \Delta_0$ for $\alpha_0 = 0.02$ at low 
temperature.
Both the curves for sequential tunneling (gray solid line) and sequential plus
cotunneling (dotted line) are temperature independent.
A full second-order perturbation theory, however, shows deviations from this
universal scaling behavior due to charging-energy gap renormalization.
The chosen temperatures are $k_B T/E_C= 0.01$ (dashed line) and $0.0001$ 
(black solid line).
We analyze in detail the reduction of slope at resonance (see inset) and the 
shift of the maximum away from resonance with decreasing temperature.
}
\label{SlowT}
\end{figure}

These features of the thermopower have been explained by Turek and 
Matveev\cite{TurekPRB02} by considering sequential plus cotunneling processes 
(the terms $g^{\text{seq}}$ and $g^{\text{cot}}$ only).
They postulate a universal low-$T$ behavior, whereby $S$ scales as 
$S(\beta \Delta_N)$. 
We find that this does not hold true for a complete higher-order calculation. 
Taking into account the renormalization of system parameters due to quantum 
fluctuations lets the thermopower deviate from universal behavior as 
illustrated in Fig.~\ref{SlowT}, in which the thermopower is plotted as
a function of $\beta \Delta_0$. 
Conversely, these deviations allow an insight into the renormalization process 
and reveal the rich physics missed by taking into account cotunneling 
processes only.

In the following, we will concentrate on two distinctive features of 
Fig.~\ref{SlowT}: On the behavior close to the resonance of sequential 
tunneling (at half integer values of $n_x$) and on the position of the 
maximum of thermopower.

\subsubsection{Reduction of charging-energy gap}

\begin{figure}[]
\centerline{\includegraphics[width=0.95\columnwidth]{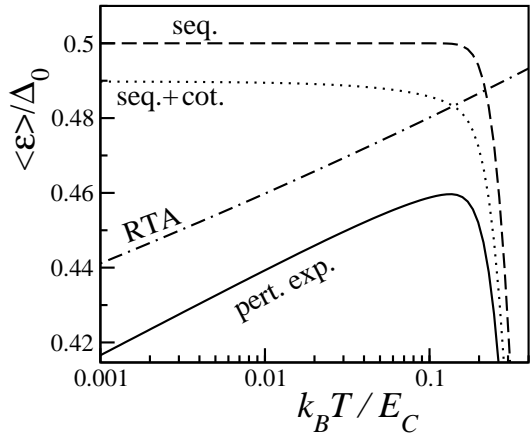}}
\caption{Average energy of the transported particles as a function of temperature for 
coupling $\alpha_0=0.01$. 
The logarithmic temperature dependence typical for Kondo-physics indicates 
a renormalization of the charging-energy gap.
We display the sequential-tunneling result (dashed line), the ``universal'' 
(temperature-independent) result of sequential and cotunneling (dotted) and 
the full next-to-leading-order perturbative result of Eq.~(\ref{Squotient})
(solid). 
RTA (dot-dashed line) gives a similar low-temperature result for the slope in 
this Figure, but the off-set is unknown.}
\label{renormfiga}
\end{figure}

Close to the resonance at $n_x=1/2$, transport is associated with charge 
excitations $0\leftrightarrow 1$, and thermopower is linear in $n_x$.
The sequential-tunnelig result $S^{\text{seq}}=- (k_B/e) \beta\Delta_0/2$ 
corresponds to an average energy $\Delta_0/2$ of the contributing lead 
electrons.
Including cotunneling processes slightly reduces the slope (dotted line) by a
factor that is independent of temperature.
The slope is further reduced when taking into account all second-order 
contributions.
Using Eq.~(\ref{onsager}) as a definition of the average energy we see that
the reduction of the slope reflects a reduction of the average energy of the
contributing lead electrons.
In Fig.~\ref{renormfiga} we display the average energy defined via 
Eq.~(\ref{onsager}) close to resonance as function of temperature 
(this corresponds to the slope at $n_x=1/2$ in Fig.~\ref{SlowT}).

Sequential tunneling gives the ratio of $1/2$ for $\left\langle \varepsilon 
\right\rangle /\Delta_0$, reflecting energy averaging as discussed above
(dashed line in Fig.~\ref{renormfiga}).
The cotunneling regularized at resonance (dotted line) yields a constant 
reduction by the small perturbation parameter $\alpha_0$:
\begin{eqnarray} 
\frac{ g^{\text{seq}}_T + g^{\text{cot}}_{T} }{ g_V^{\text{seq}} + g^{\text{cot}}_{V} }&=& \frac{\beta\Delta_0}{2}\frac{1-4\alpha_0}{1-2\alpha_0} \nonumber\\
&=& \frac{\beta\Delta_0}{2} (1-2\alpha_0) +{\text{O}}(\alpha_0^2).\label{slope_cot}
\end{eqnarray}
The full next-to-leading-order calculation (solid line) of $S$ results in
\begin{equation} \label{slope_renorm}
\left\langle \varepsilon \right\rangle /\Delta_0 = \frac{1}{2} \left[ 1 -2\alpha_0\left(2 + \gamma + \ln{\left(\frac{\beta E_C}{\pi}\right)}\right)\right],
\end{equation}
with Euler's constant $\gamma=0.5772\hdots\;$. 
The logarithmic temperature dependence directly reflects the renormalization 
of the charging-energy gap, cf.~Eq.~(\ref{Deltarenormcot}), which indicates
many-channel Kondo physics.

We also show the result from the resonant-tunneling approximation (dot-dashed 
line), which is obtained from numerical integration of Eqs.~(\ref{G_V}) and
(\ref{G_T}) with Eq.~(\ref{spec_RTA}).
The logarithmic temperature dependence again reflects the renormalization of
the level position
$\tilde \Delta_0/\Delta_0 = \left[ 1 + 2\alpha_0\left( \gamma + \ln{\left( 
\beta D/2\pi \right) }\right)\right]^{-1}$ close to resonance.
While the logarithmic temperature behavior and consequently the slope in Fig.~\ref{renormfiga} 
is reliably predicted by RTA, the absolute vertical position
depends on the choice of the high-energy cutoff (here we took $D=E_C$).

As thermopower measures the average energy of transport it yields a direct extraction of the 
renormalized charging-energy gap $\tilde\Delta$ (via the slope of thermopower at resonance). This complements in a very appealing manner 
electrical conductance measurements, which reveal the renormalization of the coupling constant $\tilde\alpha$. As discussed in Ref.~\onlinecite{KoenigSchoellerSchoenPRB98} and experimentally observed in Refs.~\onlinecite{Joyez97} and \onlinecite{Wallisser02} this renormalization of coupling is seen as logarithmic reduction of the maximal linear conductance at low temperatures.

\subsubsection{Maximum of thermopower}

\begin{figure}[b]
\centerline{\includegraphics[width=0.98\columnwidth]{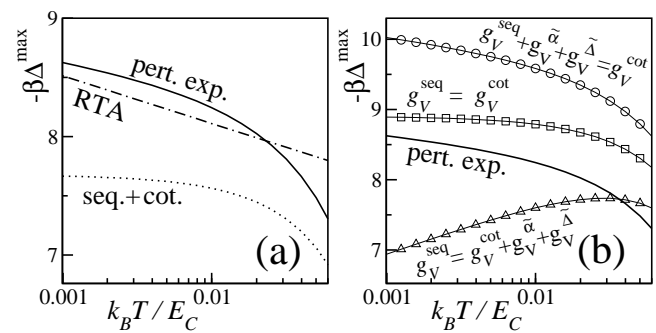}}
\caption{(a) Position of the maximum of thermopower for $\alpha_0=0.01$.
When taking only sequential and cotunneling into account, the maximum's 
position approaches a constant at low temperature (dotted line).
In the full next-to-leading-order perturbative calculation (solid line) and 
the RTA (dot-dashed line),
the maximum moves away from resonance with decreasing temperature. 
(b) Approximative determination of the temperature dependence of the 
maximum's position [thick solid line as in (a)].
The crossover between sequential and cotunneling (squares) becomes temperature
independent at low-temperature.
The correct temperature dependence is reproduced when the second-order terms
associated with renormalization of the system parameters are added to the
sequential-tunneling (circles) but not to the cotunneling (triangles) 
contribution.}
\label{renormfigb}
\end{figure}

The renormalization of the charging-energy gap not only modifies the slope of 
thermopower around $n_x$, it also shifts the position of maximum.
Figure \ref{renormfigb}(a) shows the (numerically determined) position of 
the maximum as a function of temperature.
With only sequential and cotunneling taken into account,\cite{TurekPRB02}
the maximum position approaches a constant when lowering the temperature
(dotted line) as a consequence of the prososed unversal scaling behavior.
In a full next-to-leading-order theory, however, the maximum position grows 
logarithmically with decreasing temperature (solid line).
The same low-temperature behavior is reproduced by the resonant-tunneling 
approximation (dot-dashed line).

Since the exact analytic expression for the maximum position is not transparent,
we can gain some more insight in the origin of the maximum and the temperature dependence of
its position by using the following approximate treatment.
Close to resonance, the average energy from sequential tunneling (and thus the thermopower) increases 
linearly with $\Delta_0$.
Away from resonance, the cotunneling contribution dominates, and the 
thermopower decreases proportional to $T^2/\Delta_0$.
The total thermopower including both types of processes is the average
of the individual thermopower expression, weighted with the electrical
conductances $g_V^{\text{seq}/\text{cot}}$
\begin{equation} \label{averageenergy}
-S eT= \left\langle \varepsilon \right\rangle = \frac{ g_V^{\text{seq}} \Delta_0/2 + g_V^{\text{cot}}(k_BT)^2/\Delta_0}{ g_V^{\text{seq}} + g_V^{\text{cot}} }.
\end{equation}  
Therefore, the maximum position is roughly 
at the point where sequential and cotunneling electrical conductance
coincide\cite{TurekPRB02} (we disregard any numerical factors here).

In Fig.~\ref{renormfigb}(b), we show the maximum position determined in this way.
If sequential and cotunneling processes were the only contributions to be 
considered, the obtained maximum position is of the order of the 
numerically determined value and is constant at low temperature (squares).
However, in a full second-order calculation there are additional terms, as 
discussed above.
How do these terms fit into this picture?
When only looking at the power of the perturbation expansion in $\alpha_0$
one might consider these second-order terms belonging to cotunneling.
This, however, does result in a completely wrong temperature dependence 
(triangles).
The maximum position should rather be determined by equating the
cotunneling electrical conductance with that of sequential tunneling plus
the extra second-order terms, that are interpreted as renormalization 
corrections to sequential tunneling (circles).
In this case, the correct temperature dependence is reproduced.
This, once again, supports the picture of renormalization.

The main effect of renormalization on the maximum position is that
the peak of the electrical conductance $g_V^{\text{seq}}$ around resonance is broadened since
the renormalized charging-energy gap 
$\tilde \Delta = \Delta_0 [ 1 - 2\alpha_0 ( \text{const.} + \ln \beta E_C )]$ 
is reduced, i.e., the system is moved \emph{closer} to resonance.
As a consequence, the maximum position moves \emph{away} from resonance,
$\Delta^{\text{max}} = \Delta_0 [ 1 + 2\alpha_0 ( \text{const.} + 
\ln \beta E_C )] +  {\text{O}}(\alpha_0^2)$,
to compensate for this renormalization, so that the renormalized maximum position is left unchanged. 
This is indeed the asymptotic behavior found in Fig.~\ref{renormfigb}.
In conclusion, the temperature dependence of the maximum position reflects the 
temperature-dependent renormalization of the charging-energy gap.

\section{Conclusions\label{Conclusions}}
In this paper, we investigated the low-temperature properties of thermal 
conductance and thermopower of a metallic single-electron transistor. 
We presented two approximation schemes for analyzing 
higher-order contributions associated with quantum fluctuations: 
A systematic perturbative expansion in the tunnel-coupling strength and the 
nonperturbative resonant-tunneling approximation.
Both these schemes reveal qualitatively similar physical effects and 
coincide for weak coupling. 
In particular, we find that quantum-fluctuation-induced renormalizations of 
the charging-energy gap and the tunnel-coupling strength affect the 
thermoelectric properties.
They yield logarithmic temperature dependences typical for the (many-channel) 
Kondo effect.

For the thermal conductance, renormalization of tunnel-coupling strength and
charging-energy gap results in a suppression and a broadening of the resonance 
features, respectively (see Fig.~\ref{G_Tfig}).
The effect of charging-energy-gap renormalization is most striking in the
thermopower.
It destroys the universal low-temperature scaling that would follow from
considering sequential and cotunneling processes only (Fig.~\ref{SlowT}).
The reduction of the charging-energy gap is reflected in a reduction of the 
slope of the thermopower at resonance, and in a shift of the maximum's position
away from resonance (Figs.~\ref{renormfiga} and \ref{renormfigb}, respectively).
This is due to the fact, that thermopower can be interpreted as measure of the 
average energy of transported particles.
Therefore, an experimental observation of the charging-energy gap 
renormalization in the thermopower would provide an appealing complement to 
the tunnel-coupling renormalization measured in the electrical 
conductance.~\cite{Joyez97,Wallisser02}
Certainly measuring such effects in thermopower is more challenging, as precise control of 
temperature bias and gate voltage is required. 
Sequential\cite{StaringEPL93,DzurakSSC93} and cotunneling\cite{DzurakPRB97} 
have been observed in thermopower measurements some years after these tunneling
processes had first been investigated in the electrical 
conductance.\cite{FultonDolan87,GeerligsPRL90,EilesPRL92}
In continuing this track record also the renormalization effects on thermopower
predicted in this paper may be within reach of future experiments.

\begin{acknowledgments}
We acknowledge discussions with G.~Johansson and financial support by DFG via Graduiertenkolleg 726.
\end{acknowledgments}


\end{document}